\documentclass[submission,copyright,creativecommons]{eptcs}
\providecommand{\event}{TARK 2025} % Name of the event you are submitting to

\usepackage{iftex}

\ifpdf
  \usepackage{underscore}         % Only needed if you use pdflatex.
  \usepackage[T1]{fontenc}        % Recommended with pdflatex
\else
  \usepackage{breakurl}           % Not needed if you use pdflatex only.
\fi

\usepackage{amsfonts}
\usepackage{amssymb}
\usepackage{amsmath}

\newtheorem{theorem}{Theorem}

\newtheorem{definition}{Definition}

\newtheorem{lemma}{Lemma}

\newtheorem{remark}{Remark}

\numberwithin{equation}{section}

\title{Prudent Rationalizability and the Best Rationalization Principle}
\author{Nicodemo De Vito
\institute{Department of Economics and Statistics\\
University of Salerno\\
Salerno, Italy}
\email{ndevito@unisa.it}
}
\date{June 2025}

\def\titlerunning{Prudent Rationalizability and the Best Rationalization Principle}
\def\authorrunning{N. De Vito}

\begin{document}
\maketitle

\begin{abstract}
We study cautious reasoning in finite sequential games played by agents with
perfect recall. Our contribution lies in formulating a definition of \textit{%
prudent rationalizability}\ (Heifetz et al. 2021, \textit{BEJTE}) as an
iterative reduction procedure of beliefs. To this end, we represent the
players' beliefs by systems of conditional \textit{non-standard} probability
measures. The key novelty is the notion of \textit{c-strong belief}, a
non-standard, \textquotedblleft cautious\textquotedblright\ version of
strong belief (Battigalli and Siniscalchi 2002, \textit{JET}). Our
formulation of prudent rationalizability embodies a \textquotedblleft best
rationalization principle\textquotedblright\ similar to\ the one that
underlies the solution concept of strong rationalizability. The main results
show the equivalence between the proposed definition with the one originally
put forth by Heifetz et al. (2021) in terms of conditional beliefs
represented by standard probabilities. In particular, it is shown that
prudent rationalizability can be algorithmically characterized by iterated
admissibility. Finally, our formulation can be extended to sequential games
with unawareness.

\end{abstract}

\section{Introduction}

\textit{Prudent rationalizability} is a solution concept for games where
some moves are sequential, henceforth \textit{sequential games}. It was
introduced by Heifetz et al. (2021) for the analysis of cautious reasoning
in the framework of sequential games with \textit{unawareness} (Heifetz et
al., 2013), i.e., games which allow for asymmetric awareness of the players
concerning their actions. As argued by the authors, studying cautious
reasoning is natural in such a context since, as the play unfolds, players
frequently become aware of actions they (or the co-players) could have
chosen. With this, they put forward the notion of prudent rationalizability
as a \textquotedblleft cautious\textquotedblright\ analogue of \textit{%
strong rationalizability}, aka extensive-form rationalizability (Pearce
1984, Battigalli 1997), a solution concept that captures forward-induction
thinking. In the context of limited awareness, prudent rationalizability
proved to be important in applications of interest---see Meier and Schipper
(2024) and the relevant references therein.

The aim of our work is to address an issue left open by Heifetz et al.
(2021, Section 6). Specifically, their definition of prudent
rationalizability is in terms of an iterative reduction procedure of
strategies. The open question is whether a definition via a reduction
procedure for conditional beliefs---in the style of strong
rationalizability---is feasible. Such an alternative definition would be
desirable because, as we will argue below, it would shed light on the
possible epistemic foundations for prudent rationalizability.

To address the above question, here we study sequential games played by agents with perfect recall, without assuming unawareness. However, the definitions and results can be extended to the case of games with unawareness. We will show that our formulation of prudent rationalizability
provides a positive resolution to the open issue in Heifetz et al. (2021). We will also show that such a formulation embodies a \textit{best rationalization principle}%
\ similar to\ the one that underlies the solution concept of strong rationalizability (Battigalli 1996).

We base our analysis on two key notions which pertain to the representation
of players' beliefs. First, we represent the players' beliefs by \textit{%
conditional non-standard probability systems} (CNPSs): as the play unfolds,
each players holds conditional beliefs about the co-players' strategies
which are updated or revised according to the rules of conditional
probability. More importantly, such (conditional) beliefs are represented by
probability measures whose range is a subset of the hyperreal field $^{\ast }%
\mathbb{R}$, a non-Archimedean field that includes the real numbers but also
has \textit{infinitesimals} (i.e., numbers that are positive but smaller than
than any positive number). This representation of beliefs allows to
formalize the idea that a player is both cautious (no strategy of the
co-players is ruled out) and, at the same time, he/she deems an event
infinitely more likely than another.

The other relevant notion is \textit{c-strong belief}, a \textquotedblleft
cautious\textquotedblright\ version of strong belief (Battigalli and
Siniscalchi 2002). We say that a player c-strongly believes an event $E$ if
he/she \textit{cautiously believes} it at the beginning of the game; as the
play unfolds, he/she continues to do so as long as $E$ is not falsified by
the evidence. The concept of cautious belief was introduced by Catonini and
De Vito (2024) for their epistemic analysis of iterated weak dominance in
finite static games. Loosely speaking, a player cautiously believes an event 
$E$\ if he/she deems every co-players' strategy in $E$ infinitely more
likely than the strategies in the complement not-$E$. As shown by Catonini
and De Vito (2024), this notion of belief embodies a form of caution
relative to the event.

We illustrate our approach with two results. First, we prove (Theorem 1 in
Section 6) that our formulation of prudent
rationalizability can be given an algorithmic characterization in terms of
iterated admissibility, i.e., the set of strategies surviving the maximal
iterated deletion of weakly dominated strategies. Thus, although the
conceptual definition of prudent rationalizability depends on the
extensive-form structure of the game, it has a \textquotedblleft
strategic-form\textquotedblright\ flavor from an operational standpoint
(i.e., in terms of finding out the set of strategies). Theorem 1 is, in
fact, a version---for the case of games without unawareness---of Theorem 3
in Meier and Schipper (2024), which is stated under the definition of
prudent rationalizability given by Heifetz et al. (2021). Our second result
(Theorem 2 in Section 6) shows the equivalence
between our definition of prudent rationalizability and the one put forth by
Heifetz et al. (2021). As already mentioned, this equivalence extends to
sequential games with unawareness.

The present work is part of a research program concerning the epistemic
foundations of prudent rationalizability. A paper in progress by the author
of this work shows that, by means of hypermethods---mainly, the \textit{%
transfer principle} of non-standard analysis---a \textquotedblleft
canonical\textquotedblright\ epistemic model for hierarchies of CNPSs
exists. With this, it should be possible to formally provide epistemic
conditions for prudent rationalizability: the natural starting point is
Theorem 3 in Catonini and De Vito (2024), which provides an epistemic
foundation for \textit{lexicographic rationalizability} (Stahl 1995)---a
version of prudent rationalizability for the case of static games---under
the assumption called \textquotedblleft transparency of
cautiousness.\textquotedblright \footnote{%
The following issue is quite subtle, and can be properly understood by a
reader who is familiar with Heiftez at al. (2021, Sections 4.2-4.3). As
argued by the authors, there is a tension between caution (or
\textquotedblleft prudence\textquotedblright ) and the logic of the
rationalization underlying forward-induction thinking. A strategy can be
incautious (or \textquotedblleft imprudent\textquotedblright ) but strongly
rationalizable. Heifetz et al. (2021, Section 4.2) argue that
\textquotedblleft ...the definition of prudent rationalizability resolves
this tension unequivocally in favor of the prudence
consideration.\textquotedblright\ \ In our view, the implicit assumption
underlying the above statement is that caution is \textquotedblleft
transparent\textquotedblright\ in the terminology of Catonini and De Vito
(2024) or Battigalli and De Vito (2021): whenever a history reveals that a
player's strategy is inconsistent with prudent rationalizability, such a
strategy is always interpreted by the co-players as cautious
(\textquotedblleft prudent\textquotedblright ), but it is not strongly
rationalizable. Put differently, caution is given \textbf{doxastic} (or
\textquotedblleft \textbf{epistemic}\textquotedblright ) \textbf{priority}
at histories inconsistent with the predicted path of play (cf. Battigalli
and De Vito 2021, Section 6).}

\textit{Organization}. Section 2
introduces the required mathematical notation and auxiliary notions from
non-standard analysis. Section 3 introduces the
game-theoretic framework and formalizes the players' beliefs. Section 4 defines c-strong belief. Section 5 introduces the proposed definition of prudent
rationalizability. Section 6 contains the results.
All proofs are in the Appendix.

\section{Preliminaries\label{Section Mathematical Preliminaries}}

This section reviews some basic concepts from non-standard analysis
pertaining to the elements of $^{\ast }\mathbb{R}$, the hyperreal
line---see, e.g., Goldblatt (1998) for an excellent introduction. Here, we
restrict attention to the set of non-negative hyperreal numbers less than or
equal to $1$, denoted by $^{\ast }\left[ 0,1\right] $. Specifically, $^{\ast
}\left[ 0,1\right] $\ is a non-standard extension of the set of real numbers 
$\left[ 0,1\right] $, and it contains the latter. In particular, $^{\ast }%
\left[ 0,1\right] $\ also contains positive infinitesimal numbers, i.e.,
numbers strictly greater than zero but smaller than any positive real
number. The existence of such a non-standard extension of $\left[ 0,1\right] 
$\ follows from the ultrapower construction of the hyperreal field---see
Luxemburg (1962).

Formally, a number $x\in $ $^{\ast }\mathbb{R} $ is an \textbf{%
infinitesimal} if $\left\vert x\right\vert<1/n$\ for each natural number $n\in \mathbb{N}$. The
unique infinitesimal real number is $0$. For any $x,y\in $ $^{\ast }\left[
0,1\right] $, say that $x$\ and $y$ are \textbf{infinitely close}, and write 
$x\simeq y$, if $\left\vert y-x\right\vert $ is an infinitesimal. The set of
positive infinitesimals is denoted by $\mathrm{Hal}^{+}\left( 0\right) $, i.e.,
the set of all $x\in $ $^{\ast }\left[ 0,1\right] $ which are infinitely
close to $0$.\footnote{%
In the terminology of non-standard analysis, $\mathrm{Hal}^{+}\left(
0\right) $\ is the positive part of the \textbf{monad},\ or \textbf{halo}%
,\ of zero.} The \textbf{standard part} of $x\in $ $^{\ast }\left[ 0,1\right]
$, denoted by $\mathrm{st}\left( x\right) $, is the unique real number $y$\
such that $x\simeq y$. Every $x\in $ $^{\ast }\left[ 0,1\right] $\ can be
uniquely written as $x=y+\epsilon $, where $y=\mathrm{st}\left( x\right) $\
and $\epsilon \in $ $^{\ast }\mathbb{R}$ is infinitesimal.

Given $x,y\in $ $^{\ast }\mathbb{R}^{+}$, say that $x$ is \textbf{infinitely
greater} \textbf{than} $y$ (or $y$\ is \textbf{infinitely smaller\ than}\ $x$%
) if $x>ny$ for all $n\in \mathbb{N}$. Note that if $x$ is infinitely
greater than $y$, then $x>0$. It is easy to check that $x$\ is infinitely
greater than $y$ \textit{if and only if} $\mathrm{st}\left( y/x\right) =0$.%
\footnote{%
Let $x>ny$ for each $n\in \mathbb{N}$. Then $y/x<1/n$ for each $n\in \mathbb{%
N}$, i.e., $y/x$ is infinitesimal, which yields $\mathrm{st}\left(
y/x\right) =0$. For the reverse implication, we argue by contraposition.
Suppose that $x\leq ny$ for some $n\in \mathbb{N}$. Then $y/x\geq 1/n$ for
some $n\in \mathbb{N}$, i.e., $y/x$ is not infinitesimal; hence, $\mathrm{st}%
\left( y/x\right) \neq 0$.}

Given any \textit{finite} set $X$, probability measures on $X$ are defined
from the power set of $X$\ to $^{\ast }\left[ 0,1\right] $. Such
probabilities are called non-standard. We sometimes say that a probability
measure is standard if its range is the set $\left[ 0,1\right] \subseteq 
\mathbb{R}$. We let $^{\ast }\Delta \left( X\right) $\ and $\Delta \left(
X\right) $ denote the sets of, respectively, non-standard and standard
probabilities on $X$. The latter set can be regarded as a subset of the
former, and it makes sense to write $\Delta \left( X\right) \subseteq $ $%
^{\ast }\Delta \left( X\right) $.

\section{Framework\label{Section: framework}}

We introduce the building blocks of the analysis, namely, finite games with
observed actions (Section 3.1)
and systems of conditional non-standard probabilities (Section 3.2).

\subsection{Finite Games with Observed Actions\label{Subsection: finite game
observed actions}}

Throughout, we focus on finite multistage games with perfect monitoring of
past actions.\footnote{%
The restriction to finite multistage games with observed actions is only for
the sake of notational simplicity. The techniques and results in this paper
can be extended to finite sequential games played by agents with perfect
recall.}

A \textbf{finite game with observed actions} is represented by a structure $%
G:=\left\langle I,\bar{H},\left( A_{i},u_{i}\right) _{i\in I}\right\rangle $%
\ where:

\begin{itemize}
\item $I$ is a finite set of \textbf{players}, and, for each $i\in I$, $%
A_{i} $ is a finite, nonempty set of feasible \textbf{actions}.

\item $\bar{H}$ is a finite tree of feasible\textit{\ }\textbf{histories},
that is, of sequences of action profiles $a\in A:=\times _{i\in I}A_{i}$.
The root of $\bar{H}$ is the \textbf{initial history}, or \textquotedblleft
empty sequence,\textquotedblright\ and is denoted by $\varnothing $. We let $%
Z$ denote the set of \textbf{terminal} histories (or \textbf{paths}), and $%
H:=\bar{H}\backslash Z$\ is the set of \textbf{non-terminal} histories.

\item For each $h\in H$, the set of feasible action profiles%
\begin{equation*}
A(h):=\left\{ a\in A:\left( h,a\right) \in \bar{H}\right\}
\end{equation*}%
is such that $A(h)=\times _{i\in I}A_{i}(h)$, where $A_{i}(h)$\ is the
projection of $A(h)$\ on $A_{i}$.

\item For each $i\in I$, $u_{i}:Z\rightarrow \mathbb{R}$\ is the utility
function for player $i$.
\end{itemize}

The interpretation of $G$ is that, as the game unfolds, each player is
informed of the sequence of action profiles that has just occurred.
Specifically, it is assumed more: as soon as a history $h$ occurs it becomes
common knowledge that $h$ has occurred.

For any $h\in \bar{H}$, we write $h\prec h^{\prime }$\ ($h\preceq h^{\prime
} $) if sequence/history $h$\ is a strict (weak) prefix of $h^{\prime }$,
and we say that $h$ \textbf{(weakly) precedes }$h^{\prime }$ if ($h\preceq
h^{\prime }$) $h\prec h^{\prime }$. Player $i$ is \textbf{active} at history 
$h\in H$ if he has at least two feasible actions, and he is \textbf{inactive}
otherwise.\footnote{%
Whenever player $i$ is not active at $h\in H$, one can think of the unique
element of $A_{i}\left( h\right) $ as the \textquotedblleft
action\textquotedblright\ of waiting one's turn to move.} There are
simultaneous moves given $h$ if at least two players are active at $h$. Game 
$G$ is \textbf{static} if $H=\left\{ \varnothing \right\} $.

The analysis in this paper focuses on the following derived objects. For
each $i\in I$, let $S_{i}:=\times _{h\in H}A_{i}(h)$ and $S:=\times _{i\in
I}S_{i}$. Each $s_{i}\in S_{i}$\ is a \textbf{strategy} of player $i$, i.e.,
a function $s_{i}:H\rightarrow A_{i}$ such that $s_{i}\left( h\right) \in
A_{i}(h)$\ for all $h\in H$. The set of strategies
of players other than $i$ is $S_{-i}:=\times _{j\in I\backslash \{i\}}S_{j}$.%
\footnote{%
Our notation is standard: for any profile of sets $\left( X_{i}\right)
_{i\in I}$, we let $X_{-i}:=\times _{j\in I\backslash \{i\}}X_{j}$ with
typical element $x_{-i}:=\left( x_{j}\right) _{j\neq i}$ $\in X_{-i}$.}

Each strategy profile $s:=\left( s_{i}\right) _{i\in I}\in S$\ induces a
unique terminal history $\zeta \left( s\right) $, where $\zeta :S\rightarrow
Z$ denotes the \textbf{path function}. With this, for each $h\in H$, the set
of strategies inducing $h$ is%
\begin{equation*}
S\left( h\right) :=\left\{ s\in S:h\prec \zeta \left( s\right) \right\} 
\text{.}
\end{equation*}%
The projection%
\begin{equation*}
S_{i}\left( h\right) :=\left\{ s_{i}\in S_{i}:\exists s_{-i}\in
S_{-i},\left( s_{i},s_{-i}\right) \in S\left( h\right) \right\}
\end{equation*}%
is the set of strategies of $i$ that \textbf{allow} $h$ (i.e., do not
prevent the realization of $h$).\footnote{%
If $i$ carries out strategy $s_{i}$, then $h$ can be reached; whether it is
reached depends on the play of $i$'s co-players.} Analogously,%
\begin{equation*}
S_{-i}\left( h\right) :=\left\{ s_{-i}\in S_{-i}:\exists s_{i}\in
S_{i},\left( s_{i},s_{-i}\right) \in S\left( h\right) \right\}
\end{equation*}%
is the set of strategy profiles of $i$'s co-players that allow $h$. It is
worth noting that, in a game with observed actions, $S\left( h\right)
=\times _{i\in I}S_{i}\left( h\right) $\ for each $h\in H$. Finally,%
\begin{equation*}
U_{i}:=u_{i}\circ \zeta :S\rightarrow \mathbb{R}
\end{equation*}
determines the (strategic-form) payoff $U_{i}\left( s\right) =u_{i}\left(
\zeta \left( s\right) \right) $ of player $i$ as a function of $s$.

\subsection{Conditional Beliefs\label{Subsection: CPS's}}

For conciseness, the definitions in this section apply to a game $%
G:=\left\langle I,\bar{H},\left( A_{i},u_{i}\right) _{i\in I}\right\rangle $%
. Each player's beliefs are represented by arrays of non-standard
probability measures indexed by elements of a (finite) collection of
\textquotedblleft conditioning events,\textquotedblright\ i.e., sets of the
form $S_{-i}\left( h\right) $ for $h\in H$.\ Such events can be thought of
as representing the evidence that a player can obtain in a game. For
instance, if history $h\in H$ occurs, player $i$ learns that the co-players
are behaving according to a strategy profile in $S_{-i}\left( h\right) $. If 
$i$'s belief before the realization of history $h$ assigned probability $0$
to $S_{-i}\left( h\right) $, then the occurrence of $h$ falsifies $i$'s
earlier belief, which has to be revised, rather than updated according to
the chain rule of conditional probabilities (formally introduced below).

Let $\mathcal{S}_{-i}:=\left\{ F\subseteq S_{-i}:\exists h\in
H,F=S_{-i}\left( h\right) \right\} $\ denote the collection of conditioning
(or \textquotedblleft observable\textquotedblright ) events about the
behavior of $i$'s co-players. We also use the following notation: for any 
\textit{finite} set $X$\ and \textbf{event} (subset) $E\subseteq X$, let $%
^{\ast }\Delta \left( E\right) :=\left\{ \mu \in \text{ }^{\ast }\Delta
\left( X\right) :\mu \left( E\right) =1\right\} $\ be the set of
non-standard probability measures that assign probability $1$ to $E$. As for
standard probabilities, the set $\Delta \left( E\right) $\ is defined
accordingly.

\begin{definition}
An array of probability measures $\mu :=\left( \mu (\cdot |C)\right) _{C\in 
\mathcal{S}_{-i}}$ is a \textbf{conditional non-standard probability system}
(\textbf{CNPS}) for player $i\in I$\ if

(i) $\mu (\cdot |C)\in $ $^{\ast }\Delta \left( C\right) $ for all $C\in 
\mathcal{S}_{-i}$, and

(ii) for all $C,D\in \mathcal{S}_{-i}$ and $E\subseteq S_{-i}$,%
\begin{equation}
E\subseteq D\subseteq C\Rightarrow \mu (E|C)=\mu (E|D)\mu (D|C)\text{.}
\label{ChainRuleCPS}
\end{equation}%
The array $\mu $ is \textbf{conditional probability system} (\textbf{CPS})
if it is a CNPS such that $\mu (\cdot |C)\in $ $\Delta \left( C\right) $\
for all $C\in \mathcal{S}_{-i}$.
\end{definition}

Condition (3.1) is the \textbf{chain\ rule }of conditional
probabilities, and it can be written as follows: if $E\subseteq D\subseteq C$%
, then%
\begin{equation*}
\mu (D|C)>0\Rightarrow \mu (E|D)=\frac{\mu (E|C)}{\mu (D|C)}\text{.}
\end{equation*}%
We let $\Delta ^{\mathcal{S}_{-i}}\left( S_{-i}\right) $\ and $^{\ast
}\Delta ^{\mathcal{S}_{-i}}\left( S_{-i}\right) $\ denote, respectively, the
sets of CPSs and CNPSs for player $i\in I$. As argued above, the former set
can be regarded as a subset of the latter, so $\Delta ^{\mathcal{S}%
_{-i}}\left( S_{-i}\right) \subseteq $ $^{\ast }\Delta ^{\mathcal{S}%
_{-i}}\left( S_{-i}\right) $. Throughout this paper, we will often use the
notation $\mu :=\left( \mu \left( \cdot \left\vert S_{-i}\left( h\right)
\right. \right) \right) _{h\in H}$\ to refer to an arbitrary element of $%
^{\ast }\Delta ^{\mathcal{S}_{-i}}\left( S_{-i}\right) $\ (or $\Delta ^{%
\mathcal{S}_{-i}}\left( S_{-i}\right) $).

CPSs are used as representations of players' beliefs to model sophisticated
strategic thinking in sequential games, such as forward-induction thinking
(Battigalli and Siniscalchi 2002). However, as we will see below (Sections %
4-6), there is a sense
in which CNPSs are more expressive than CPSs: unlike a CPS, a CNPS also
allows to properly formalize forms of strategic thinking based on notions of
\textquotedblleft caution.\textquotedblright

We point out that a CNPS can be viewed as a non-standard analogue of a 
\textit{system of conditional lexicographic probabilities} (SCLP), a concept
introduced by Asheim and Perea (2005) in their analysis on cautious
reasoning in sequential games. In contrast to a CNPS, an SCLP represents
player $i$'s beliefs---conditional on every $C\in \mathcal{S}_{-i}$---by
means of \textit{lexicographic probability systems}, i.e., finite sequences
of probability measures, where the different measures are given in a
decreasing order of importance. In other words, and using the notation of
the current paper, each conditional belief $\mu (\cdot |C)$\ for player $i$
is a lexicographic probability system. The axiomatic foundation of
non-Archimedean expected utility in Blume et al. (1991)\ shows that
non-standard probabilities and lexicographic probability systems are
equivalent in terms of representations of players' preferences.

That said, it makes sense to consider SCLPs as representations of players'
conditional beliefs.\footnote{%
This alternative option was suggested by Heifetz et al. (2001, Section 6)
for the formulation of prudent rationalizability as an iterative reduction
procedure for conditional beliefs. Some belief operators defined in Section 4 in terms of non-standard probabilities can be
equivalently described in terms of lexicographic probability systems---see
Catonini and De Vito (2020, 2024). The alternative characterization in terms
of SCLPs makes the analysis more complex, but it would not change the
substance of our results.} However, we represent players' beliefs as CNPSs
since this representation allows a clear-cut comparison between prudent
rationalizability and other solution concepts, such as strong
rationalizability.

\section{c-Strong Belief\label{Section c strong belief}}

Here, we define the concept of \textit{c-strong belief}, which will be used
for the formulation of prudent rationalizability. All definitions in this
section apply to a given game $G:=\left\langle I,\bar{H},\left(
A_{i},u_{i}\right) _{i\in I}\right\rangle $. For the reader's convenience,
it may be useful to consult Section 2
for the mathematical formalism used here.

Fix a CNPS $\mu :=\left( \mu \left( \cdot \left\vert S_{-i}\left( h\right)
\right. \right) \right) _{h\in H}$\ for a player $i\in I$ and fix an event $%
E\subseteq S_{-i}$. Intuitively, we say that $E$ is c-strongly believed
under $\mu $\ if, for every history $h$ consistent with $E$, every strategy
profile $s_{-i}$ in $E$\ is deemed infinitely more likely than every
strategy profile in not-$E$ (i.e., the complement of $E$). So, to define
c-strong belief, we first need to understand the idea of \textquotedblleft
infinitely more likely than.\textquotedblright

Formally, fix a history $h\in H$ and events $E,F\subseteq S_{-i}\left(
h\right) $\ such that $E\cap F\neq \emptyset $. We say that $E$ is \textbf{%
infinitely more likely than} $F$\ under $\mu \left( \cdot \left\vert
S_{-i}\left( h\right) \right. \right) $\ if%
\begin{equation*}
\mathrm{st}\left( \frac{\mu \left( F\left\vert S_{-i}\left( h\right) \right.
\right) }{\mu \left( E\left\vert S_{-i}\left( h\right) \right. \right) }%
\right) =0\text{.}
\end{equation*}%
Put differently, this definition says that $E$ is infinitely more likely
than $F$\ if $\mu \left( E\left\vert S_{-i}\left( h\right) \right. \right) $%
\ is \textit{infinitely greater than} $\mu \left( F\left\vert S_{-i}\left(
h\right) \right. \right) $.

Lo (1999) provided a preference-based definition of \textquotedblleft
infinitely more likely than.\textquotedblright\ As shown in Catonini and De
Vito (2020, 2024), Lo's notion admits a characterization in terms of
non-standard probabilities which is \textit{equivalent} to the above
definition of \textquotedblleft infinitely more likely
than.\textquotedblright\ Lo's notion is also used---at least implicitly---in
other papers, such as Meier and Perea (2024).\footnote{%
Meier and Perea (2024) formalize the idea of \textquotedblleft infinitely
more likely than\textquotedblright\ by means of an \textquotedblleft
infinitely smaller size than\textquotedblright\ relation between
non-standard numbers. It can be easily shown that such a relation can be
given a characterization in terms of the \textquotedblleft infinitely
greater than\textquotedblright\ relation---hence, the definition of
\textquotedblleft infinitely more likely than\textquotedblright\ in Meier
and Perea (2024) is equivalent to ours. Unlike us, Meier and Perea (2024)
use a non-Archimedean extension of the real line that is different from the
hyperreal line. Conceptually, this difference is immaterial: our definitions
and results still hold under the assumption that the non-standard
probability measures take value in any non-Archimedean extension of the real
field.} Similar notions of \textquotedblleft infinitely more likely
than\textquotedblright\ can be found in the literature concerning the
epistemic foundations for iterated admissibility; see Brandenburger et al.
(2008), Lee (2016).

\begin{definition}
\label{Definition c strong belief}Fix a player $i\in I$, a CNPS $\mu
:=\left( \mu \left( \cdot \left\vert S_{-i}\left( h\right) \right. \right)
\right) _{h\in H}$\ and an event $E\subseteq S_{-i}$.

(i) $E$\ is \textbf{cautiously believed} under $\mu \left( \cdot \left\vert
S_{-i}\left( h\right) \right. \right) $\ ($h\in H$) if $\mathrm{st}\left( 
\frac{\mu \left( S_{-i}\backslash E\left\vert S_{-i}\left( h\right) \right.
\right) }{\mu \left( s_{-i}\left\vert S_{-i}\left( h\right) \right. \right) }%
\right) =0$\ for all $s_{-i}\in E\cap S_{-i}\left( h\right) $.

(ii) $E$\ is \textbf{c-strongly believed} under $\mu $ if, for all $h\in H$\
such that $E\cap S_{-i}\left( h\right) \neq \emptyset $, event $E$\ is
cautiously believed under $\mu \left( \cdot \left\vert S_{-i}\left( h\right)
\right. \right) $.
\end{definition}

Before discussing the notions in Definition 2, we point out that c-strong belief admits an equivalent but more cumbersome
definition.

\begin{remark}
Since $\mu \left( E\left\vert S_{-i}\left( h\right) \right. \right) =\mu
\left( E\cap S_{-i}\left( h\right) \left\vert S_{-i}\left( h\right) \right.
\right) $ for all $h\in H$, event $E$\ is c-strongly believed under $\mu $\
if and only if, for all $h\in H$\ such that $E\cap S_{-i}\left( h\right)
\neq \emptyset $, and for all $s_{-i}\in E\cap S_{-i}\left( h\right) $,%
\begin{equation*}
\mathrm{st}\left( \frac{\mu \left( \left( S_{-i}\backslash E\right) \cap
S_{-i}\left( h\right) \left\vert S_{-i}\left( h\right) \right. \right) }{\mu
\left( s_{-i}\left\vert S_{-i}\left( h\right) \right. \right) }\right) =0%
\text{.}
\end{equation*}
\end{remark}

Cautious belief was introduced by Catonini and De Vito (2024) for the
analysis of iterated admissibility in finite static games. It requires that 
\textit{every} strategy in $E$ be deemed infinitely more likely than (every
strategy in) not-$E$. In particular, it embodies a \textit{cautious}
attitude of player $i$ towards the \textquotedblleft
believed\textquotedblright\ event.

Indeed, cautious belief requires that: (1) event $E$ is deemed infinitely
more likely than not-$E$, and (2) before entertaining the possibility that $%
E $ does not occur, player $i$ takes into account all the possible
payoff-relevant consequences (i.e., the strategy profile $s_{-i}$) of $E$.
Condition (1) corresponds to the notion of \textbf{weak belief} (Halpern
2010, Catonini and De Vito 2020), namely $\mathrm{st}\left( \mu \left(
E\left\vert S_{-i}\left( h\right) \right. \right) \right) =1$. Condition (2)
says that the player is cautious towards the (weakly) believed event $E$:
Before considering not-$E$, he/she takes into account all the possible
consequences of $E$. Thus, cautious belief in $E$ is stronger than weak
belief as it captures caution relative to $E$.

\begin{remark}
\label{Remark full support CNPS cSB}To further see how cautious belief
embodies a form of caution (as suggested by the terminology) relative to $E$%
, it could be useful to consider the special case when $E=S_{-i}$. With
this, at the initial history $\varnothing $ we have $\mu \left(
s_{-i}\left\vert S_{-i}\right. \right) >0$\ for all $s_{-i}\in S_{-i}$,
i.e., every strategy profile of $i$'s co-players is deemed possible. By the
chain rule, $\mu \left( s_{-i}\left\vert S_{-i}\left( h\right) \right.
\right) >0$\ for all $h\in H$\ and $s_{-i}\in S_{-i}\left( h\right) $.
\end{remark}

That being said, the intended interpretation of c-strong belief is simple:
Player $i$ c-strongly believes event $E$ if he/she cautiously believes it at
the beginning of the game; as the play unfolds, he/she continues to do so as
long as $E$ is not falsified by the evidence---that is, unless a history $h$
such that $E\cap S_{-i}\left( h\right) =\emptyset $ occurs.

For any event $E\subseteq S_{-i}$, let $\mathrm{cSB}_{i}\left( E\right) $\
denote the set of CNPSs of player $i$ under which $E$\ is c-strongly
believed. The reason for the terminology \textquotedblleft c-strong
belief\textquotedblright\ is clarified below. Before doing this, we find it
convenient to single out the following fact, which will be useful in the
next section.

\begin{remark}
\label{Remark nonmonotonocity cSB}c-Strong belief is not monotone: for all
events $E,F\subseteq S_{-i}$, if $E\subseteq F$, then it is not the case
that $\mathrm{cSB}_{i}\left( E\right) \subseteq \mathrm{cSB}_{i}\left(
F\right) $.
\end{remark}

In particular, non-monotonicity of c-strong belief follows from
non-monotonicity of cautious belief: as shown in Catonini and De Vito
(2024), the reason why cautious belief fails monotonicity is that player $i$
may not have towards $F$ the same cautious attitude that he has towards $%
E\subseteq F$. In particular, there could be some strategies in $F\backslash
E$\ which are not deemed infinitely more likely than not-$F$ (see Section
4.2 in Catonini and De Vito, 2024, for an example).

The notion of c-strong belief can be viewed as a \textquotedblleft cautious
analogue\textquotedblright\ of strong belief, a belief modality put forth by
Battigalli and Siniscalchi (2002) for the epistemic analysis of
forward-induction reasoning in sequential games. Strong belief is defined in
terms of CPSs, but it can also be defined in terms of CNPSs: an event $E$ is 
\textbf{strongly believed} under $\mu $ if $\mu \left( E\left\vert
S_{-i}\left( h\right) \right. \right) =1$ for all $h\in H$\ such that $E\cap
S_{-i}\left( h\right) \neq \emptyset $. Unlike c-strong belief, the notion
of strong belief does not embody caution. However, like c-strong belief,
also strong belief does not satisfy monotonicity (see Battigalli and
Siniscalchi 2002, Section 3.2).

\section{Solution Procedure\label{Section prudent rationalizability}}

In the remainder of this section, we fix a game $G:=\left\langle I,\bar{H}%
,\left( A_{i},u_{i}\right) _{i\in I}\right\rangle $. To introduce our
proposed definition of prudent rationalizability, we first need a notion of
optimality for a strategy, given a CNPS.

Fix a player $i\in I$. For each $s_{i}\in S_{i}$ and $h\in H$, let $%
s_{i}^{h} $ denote the minimal modification of $s_{i}$ allowing $h$. That
is, $s_{i}^{h}$\ is the \textquotedblleft $h$-replacement\textquotedblright\
strategy that, at every $h^{\prime }\prec h$, chooses the unique action $%
\alpha _{i}(h^{\prime },h)\in A_{i}\left( h^{\prime }\right) $ leading from $%
h^{\prime }$ toward $h$, and coincides with $s_{i}$ at all other histories.
For any CNPS $\mu _{i}$, we let $\rho _{i}\left( \mu _{i}\right) $ denote
the set of all \textbf{sequential best replies} to $\mu _{i}$, that is,%
\begin{equation*}
\rho _{i}\left( \mu _{i}\right) :=\left\{ s_{i}\in S_{i}:\forall h\in
H,s_{i}^{h}\in \arg \max_{r_{i}\in S_{i}\left( h\right) }\mathbb{E}_{\mu
_{i}}\left( U_{i}\left( r_{i},\cdot \right) |h\right) \right\} \text{,}
\end{equation*}%
where%
\begin{equation*}
\mathbb{E}_{\mu _{i}}\left( U_{i}\left( r_{i},\cdot \right) |h\right)
:=\sum_{s_{-i}\in S_{-i}\left( h\right) }U_{i}\left( r_{i},s_{-i}\right)
\mu _{i}\left( s_{-i}|S_{-i}\left( h\right) \right)
\end{equation*}%
is the (conditional) expected payoff of $r_{i}$ given $\mu _{i}\left( \cdot
|S_{-i}\left( h\right) \right) $. With this, we say that a strategy $%
s_{i}\in S_{i}$\ is \textbf{justified by} $\mu _{i}$\ (or $\mu _{i}$\ 
\textbf{justifies} $s_{i}$) if $s_{i}\in \rho _{i}\left( \mu _{i}\right) $.

It should be noted that this notion of sequential best reply (given some $%
\mu _{i}$) requires that a strategy $s_{i}$ be optimal at \textit{all}
histories, including those that are precluded by the strategy itself. In
Section 6, we will consider an alternative
notion---often called\textit{\ weak sequential optimality}---which requires
that $s_{i}$ be optimal only at histories that $s_{i}$ allows---see
Battigalli et al. (2024, Chapter 10).

\begin{definition}
\label{Definition: prudent rationalizability}Consider the following
procedure.

\begin{description}
\item[(Step $0$)] For every $i\in I$, let $S_{i}^{0}:=S_{i}$. Also, let $%
S_{-i}^{0}:=\times _{j\in I\backslash \{i\}}S_{j}$\ and $S^{0}:=\times
_{i\in I}S_{i}^{0}$.

\item[(Step $n>0$)] For every $i\in I$, let%
\begin{equation*}
S_{i}^{n}:=\left\{ s_{i}\in S_{i}:\exists \mu _{i}\in \cap_{m=0}^{n-1}%
\mathrm{cSB}_{i}(S_{-i}^{m}),s_{i}\in \rho _{i}(\mu _{i})\right\} \text{.}
\end{equation*}%
Also, let $S_{-i}^{n}:=\times _{j\in I\backslash \{i\}}S_{j}^{n}$\ and $%
S^{n}:=\times _{i\in I}S_{i}^{n}$.
\end{description}

Finally, let $S_{i}^{\infty }:=\cap_{n=0}^{\infty }S_{i}^{n}$ and $%
S^{\infty }:=\times _{i\in I}S_{i}^{\infty }$.
\end{definition}

Clearly, $\left( S^{n}\right) _{n\geq 0}$\ is a decreasing sequence, and, as
it can be shown by standard arguments, $S^{\infty }\neq \emptyset $. In
particular, since the game is finite, the sequence becomes constant after
some finite number $N\geq 0$ of steps.

Note that, in the recursive step, the procedure requires each player $i$ to
c-strongly believe all the events $S_{-i}^{m}$ ($m=0,...,n-1$), not just $%
S_{-i}^{n-1}$. \textit{Technically}, if we were to impose only the notion of
c-strong belief in $S_{-i}^{n-1}$, then $\mathrm{cSB}_{i}(S_{-i}^{n-1})$
would not necessarily be a subset of $\mathrm{cSB}_{i}(S_{-i}^{n-2})$
because, by Remark 3, c-strong belief is not
monotone. Consequently, we would not obtain a well-defined elimination
procedure---i.e., $\left( S^{n}\right) _{n\geq 0}$\ would not be decreasing.

That being said, we illustrate in some detail the \textit{conceptual}
features of the procedure; for simplicity, we focus on steps $1$-$3$.
Moreover, we will sometimes refer to the procedure as \textquotedblleft
prudent rationalizability.\textquotedblright\ The reason for using such a
terminology will be justified by the results in Section 6.

At step $1$, player $i$ plays a strategy $s_{i}$\ that is justified by some $%
\mu _{i}$ such that $\mu _{i}\left( s_{-i}\left\vert S_{-i}\left( h\right)
\right. \right) >0$\ for all $h\in H$\ and $s_{-i}\in S_{-i}\left( h\right) $%
: to see this, note that $\mathrm{cSB}_{i}(S_{-i}^{0})=\mathrm{cSB}%
_{i}(S_{-i})$ and refer to Remark 2. Thus,
besides the notion of optimality for strategies,\ the procedure requires a
form of caution (or \textquotedblleft prudence\textquotedblright ) for the
justifying CNPSs, in the sense that the players consider all strategy
profiles of the co-players possible. In what follows, we will \textit{%
informally} refer to this combination of optimality and caution as
\textquotedblleft cautious rationality.\textquotedblright

Step $2$ of the procedure requires that, on top of being cautiously
rational, all the players c-strongly believe that the co-players are
cautiously rational as well: formally, for each $i\in I$, the justifying
conditional beliefs must belong to the set $\mathrm{cSB}_{i}(S_{-i}^{1})$.

Step $3$ allows to point out an important feature of the form of strategic
thinking captured by prudent rationalizability. If player $i$ c-strongly
believes the events $S_{-i}^{1}$\ and $S_{-i}^{2}$, the he/she c-strongly
believes that the co-players' behavior is consistent with a certain degree
of strategic sophistication---namely, cautious rationality and c-strong
belief in others' cautious rationality. Suppose now that a history $h\in H$
contradicts such a degree of sophistication, that is, $S_{-i}^{2}\cap
S_{-i}\left( h\right) =\emptyset $. In this case c-strong belief in $%
S_{-i}^{2}$ leaves $i$'s beliefs at $h$ unrestricted, so $i$ may well deem
the co-players' strategies that are not cautiously rational infinitely more
likely than those consistent with $-i$'s cautious rationality; in
particular, this can happen even if $S_{-i}^{1}\cap S_{-i}\left( h\right)
\neq \emptyset $.\footnote{%
Specifically, this can occur when $S_{-i}^{2}$\ is a \textit{strict} subset
of $S_{-i}^{1}$.} However, c-strong belief in both $S_{-i}^{1}$\ and $%
S_{-i}^{2}$ rules out such a situation: whenever $S_{-i}^{1}\cap
S_{-i}\left( h\right) \neq \emptyset $, the co-players' \textquotedblleft
irrational\textquotedblright\ strategies (i.e., those in $S_{-i}\backslash
S_{-i}^{1}$)\ are deemed by $i$\ infinitely less likely than those
consistent with cautious rationality.

The above argument illustrates that prudent rationalizability---as per
Definition 3---relies on the 
\textbf{best rationalization principle} (Battigalli 1996):\ players always
ascribe to their co-players the highest degree of strategic sophistication
consistent with their past behavior, even when surprised by co-players'
behavior.\footnote{%
The best rationalization principle was originally formulated with reference
to a particular form of sophisticated strategic thinking---namely
forward-induction thinking---that underlies the solution concept of strong
rationalizability. However, the principle is stated in abstract terms; see
Battigalli (1996, p. 179). In our view, in agreement with Pierpaolo
Battigalli, it makes sense to refer to the best rationalization principle
also for the case of prudent rationalizability.} The best rationalization
principle is captured by c-strong belief in each event of the sequence $%
\left( S_{-i}^{n}\right) _{n\geq 0}$. Specifically, letting $m\left(
h\right) $\ denote the highest degree of strategic sophistication consistent
with $h\in H$,\footnote{%
Formally, $m\left( h\right) :=\max \left\{ m\in \left\{ 0,1,...,N\right\}
:S_{-i}^{m}\cap S_{-i}\left( h\right) \neq \emptyset \right\} $, where $%
N\geq 0$ is the number of steps after which the sequence $\left(
S_{-i}^{n}\right) _{n\geq 0}$ becomes constant.} we have that $\mu _{i}\in
\cap _{m=0}^{\infty }\mathrm{cSB}_{i}(S_{-i}^{m})$\ implies $\mu _{i}\in
\cap _{m=0}^{m\left( h\right) }\mathrm{cSB}_{i}(S_{-i}^{m})$ for all $h\in H$%
.

Finally, it is noteworthy that if game $G$ is static (i.e., $H=\left\{
\varnothing \right\} $), then c-strong belief corresponds to (unconditional)
cautious belief. In this case, the procedure in Definition 3 is a non-standard version of \textbf{%
lexicographic rationalizability} (Stahl 1995), a solution concept based on
lexicographic probabilities as representations of players' beliefs. Indeed,
as shown in De Vito (2023), the definition of lexicographic
rationalizability relies on the characterization of cautious belief in terms
of lexicographic probability systems (cf. Catonini and De Vito 2024).

\section{Results\label{Section main results}}

We state the main results of this paper, whose proofs can be found in the
appendix. To this end, we first need a formal definition of iteratively
admissible strategies for a game. For conciseness, the definitions and
results in this section apply to a fixed game $G:=\left\langle I,\bar{H}%
,\left( A_{i},u_{i}\right) _{i\in I}\right\rangle $.

Let $\mathcal{Q}$ be the collection of all subsets of $S$ with the form $%
Q:=\times _{i\in I}Q_{i}$, where $Q_{i}\subseteq S_{i}$ for every $i\in I$.
In what follows, for any $\sigma _{i}\in \Delta (S_{i})$, let $U_{i}(\sigma
_{i},s_{-i}):=\sum_{r_{i}\in S_{i}}\sigma _{i}\left( r_{i}\right)
U_{i}(r_{i},s_{-i})$.

\begin{definition}
Fix a set $Q\in \mathcal{Q}$. A strategy $s_{i}\in Q_{i}$\ is \textbf{weakly
dominated with respect} \textbf{to} $Q$\ if there exists a mixed strategy $%
\sigma _{i}\in \Delta (S_{i})$, with $\sigma _{i}\left( Q_{i}\right) =1$,
such that $U_{i}(\sigma _{i},s_{-i})\geq U_{i}(s_{i},s_{-i})$\ for every $%
s_{-i}\in Q_{-i}$ and $U_{i}(\sigma _{i},s_{-i}^{\prime
})>U_{i}(s_{i},s_{-i}^{\prime })$\ for some $s_{-i}^{\prime }\in Q_{-i}$.
Otherwise, say $s_{i}$\ is \textbf{admissible with respect} \textbf{to} $Q$.

If $s_{i}\in S_{i}$\ is \textit{weakly dominated (resp. admissible)} with
respect to $S$, say $s_{i}$ is \textbf{weakly dominated} (resp. \textbf{%
admissible}).
\end{definition}

The set of iteratively admissible strategies is defined recursively.

\begin{definition}
For every $i\in I$, let $\hat{S}_{i}^{0}:=S_{i}$, and for every $n\geq 1$,
let $\hat{S}_{i}^{n}$\ be the set of all $s_{i}\in \hat{S}_{i}^{n-1}$\ that
are admissible with respect to $\hat{S}^{n-1}:=\times _{j\in I}\hat{S}%
_{j}^{n-1}$. A strategy $s_{i}\in \hat{S}_{i}^{n}$\ is called $n$-\textbf{%
admissible}. A strategy $s_{i}\in \hat{S}_{i}^{\infty
}:=\cap_{n=0}^{\infty }\hat{S}_{i}^{n}$\ is called \textbf{iteratively
admissible}.
\end{definition}

By finiteness of the game, it follows from standard arguments that iterated
admissibility is a non-empty solution procedure: $\hat{S}^{\infty }:=\times
_{i\in I}\hat{S}_{i}^{\infty }\neq \emptyset $.

Let $\mathbb{N}_{0}:=\mathbb{N\cup }\left\{ 0\right\} $, i.e., $\mathbb{N}%
_{0}$ is the set of natural number including zero. The first main result of
this paper is the following theorem.

\begin{theorem}
\label{Theorem PR equivalent to IA}For all $n\in \mathbb{N}_{0}$,%
\begin{equation*}
S^{n}=\hat{S}^{n}\text{.}
\end{equation*}
\end{theorem}

We now consider the following procedure, which corresponds---for the special
case of sequential games without unawareness---to the notion of \textbf{%
prudent rationalizability} in Heifetz et al. (2021) and Meier and Schipper
(2024). Unlike Definition 3, the
following definition requires that the players' conditional beliefs be
represented by CPSs. Let $\mathrm{supp}\nu $ denote the support of a
(standard) measure $\nu \in \Delta \left( X\right) $.

\begin{definition}
\label{Definition prudent rat HMS}Consider the following procedure.

\begin{description}
\item[(Step $0$)] For every $i\in I$, let $\bar{S}_{i}^{0}:=S_{i}$. Also,
let $\bar{S}_{-i}^{0}:=\times _{j\in I\backslash \{i\}}\bar{S}_{j}^{0}$ and $%
\bar{S}^{0}:=\times _{i\in I}\bar{S}_{i}^{0}$.

\item[(Step $n>0$)] For every $i\in I$ and every $s_{i}\in S_{i}$, let $%
s_{i}\in \bar{S}_{i}^{n}$\ if and only if $s_{i}\in \bar{S}_{i}^{n-1}$\ and
there exists $\mu _{i}\in \Delta ^{\mathcal{S}_{-i}}\left( S_{-i}\right) $\
such that

\begin{enumerate}
\item $s_{i}\in \rho _{i}\left( \mu _{i}\right) $, and

\item for every $h\in H$,%
\begin{equation*}
\bar{S}_{-i}^{n-1}\cap S_{-i}\left( h\right) \neq \emptyset \Rightarrow 
\mathrm{supp}\mu _{i}\left( \cdot |S_{-i}\left( h\right) \right) =\bar{S}%
_{-i}^{n-1}\cap S_{-i}\left( h\right) \text{.}
\end{equation*}
\end{enumerate}

\item Also, let $\bar{S}_{-i}^{n}:=\times _{j\in I\backslash \{i\}}\bar{S}%
_{j}^{n}$ and $\bar{S}^{n}:=\times _{i\in I}\bar{S}_{i}^{n}$.
\end{description}

Finally, let $\bar{S}_{i}^{\infty }:=\cap _{n=0}^{\infty }\bar{S}_{i}^{n}$
and $\bar{S}^{\infty }:=\times _{i\in I}\bar{S}_{i}^{\infty }$.
\end{definition}

Before stating the second result, we find it useful to compare the procedure
described in Definition 3 to the
notion of prudent rationalizability as per Definition 6. First note that Definition 3 involves an iterative reduction procedure for
conditional beliefs: at the recursive step $n$, the set of CNPSs justifying
the strategies in $S_{i}^{n}$\ is included in the set of CNPSs justifying
the strategies in $S_{i}^{n-1}$. This in turn implies that $%
S_{i}^{n}\subseteq S_{i}^{n-1}$.

By way of contrast, Definition 6 involves an
iterative deletion of strategies: at the recursive step, it is \textit{%
required} that $\bar{S}_{i}^{n}\subseteq \bar{S}_{i}^{n-1}$. However, the
set of CPSs (not to be confused with CNPSs!) justifying the strategies in $%
\bar{S}_{i}^{n}$\ \textit{may not be} included in the set of CPSs justifying
the strategies in $\bar{S}_{i}^{n-1}$. To understand this key point, fix a
player $i\in I$\ and suppose that, at step $n+1$\ of the procedure, $\bar{S}%
_{-i}^{n}\subset \bar{S}_{-i}^{n-1}$, where the symbol $\subset $\ denotes
the \textit{strict} inclusion. Consider a strategy $s_{i}\in \bar{S}%
_{i}^{n+1}$\ and a CPS $\mu _{i}$ justifying it. Moreover, consider a
history $h$\ consistent with $\bar{S}_{-i}^{n}$, e.g., $h=\varnothing $. In
such a case, $\mu _{i}$\ must satisfy $\mathrm{supp}\mu _{i}\left( \cdot
|S_{-i}\left( \varnothing \right) \right) =\bar{S}_{-i}^{n}\cap S_{-i}\left(
\varnothing \right) =\bar{S}_{-i}^{n}$. Yet, $\mu _{i}$\ does \textit{not}
justify $s_{i}$\ at step $n$, since $\mathrm{supp}\mu _{i}\left( \cdot
|S_{-i}\left( \varnothing \right) \right) \subset \bar{S}_{-i}^{n-1}$.

Despite the above differences, the following result shows the equivalence
between the procedures in Definition 3 and Definition 6.

\begin{theorem}
\label{Theorem PR second result}For all $n\in \mathbb{N}_{0}$,%
\begin{equation*}
S^{n}=\bar{S}^{n}\text{.}
\end{equation*}
\end{theorem}

The proof of Theorem 2 is simple: it first
shows by induction the equivalence between prudent rationalizability and
iterated admissibility, viz. $\bar{S}^{n}=\hat{S}^{n}$, which is a known
result; with this, the statement follows from Theorem 1.

We conclude this section with two remarks. First, the result in Theorem 2 suggests that, at the recursive step of
Definition 6, the \textquotedblleft
full-support condition\textquotedblright\ for a justifying CPS $\mu _{i}$
can be thought of as a \textquotedblleft standard\textquotedblright\
approximation of the notion of cautious belief at history $h$: to capture
the idea that $\bar{S}_{-i}^{n-1}\cap S_{-i}\left( h\right) $\ is\
cautiously believed under\ the \textit{standard} measure $\mu _{i}\left(
\cdot |S_{-i}\left( h\right) \right) $, we require that $\mu _{i}\left(
\cdot |S_{-i}\left( h\right) \right) $\ must assign positive probability
only to the profiles in $\bar{S}_{-i}^{n-1}\cap S_{-i}\left( h\right) $.
Indeed, under a real-valued measure, an event is deemed infinitely more
likely than another only if the latter is assigned zero probability.

Second, we point out that Theorems 1 and 2 admit analogues in terms of \textquotedblleft
behaviorally equivalent\textquotedblright\ strategies (cf. Battigalli and De
Vito 2021, Section 7). To clarify, fix a player $i\in I$, and let%
\begin{equation*}
H_{i}\left( s_{i}\right) :=\left\{ h\in H:s_{i}\in S_{i}\left( h\right)
\right\}
\end{equation*}%
denote the set of non-terminal histories allowed by strategy $s_{i}$. Say
that strategies $s_{i}^{\prime }$ and $s_{i}^{\prime \prime }$ are \textit{%
behaviorally equivalent} if $H_{i}\left( s_{i}^{\prime }\right) =H_{i}\left(
s_{i}^{\prime \prime }\right) $ and $s_{i}^{\prime }\left( h\right)
=s_{i}^{\prime \prime }\left( h\right) $ for each $h\in H_{i}\left(
s_{i}^{\prime }\right) $. It can be shown by standard arguments---see
Battigalli et al. (2024, Chapter 9)---that $s_{i}^{\prime }$ and $%
s_{i}^{\prime \prime }$ are behaviorally equivalent if and only if they are 
\textit{realization equivalent}, that is, $\zeta \left( s_{i}^{\prime
},s_{-i}\right) =\zeta \left( s_{i}^{\prime \prime },s_{-i}\right) $ for all 
$s_{-i}$, which means that they induce the same terminal histories and are
observationally indistinguishable. A \textbf{reduced strategy} is an element
of the partition of $S_{i}$ induced by the behavioral equivalence relation.%
\footnote{%
Often reduced strategies are called \textquotedblleft plans of
action,\textquotedblright\ suggesting that how a strategy $s_{i}$ is defined
at histories in $H\backslash H_{i}\left( s_{i}\right) $ is irrelevant for
planning.}

With this, redefine the procedures $\left( S^{n}\right) _{n\in \mathbb{N}}$
and $\left( \bar{S}^{n}\right) _{n\in \mathbb{N}}$\ in, respectively,
Definition 3 and Definition 6 with the best-reply correspondence $\rho
_{i}\left( \cdot \right) $ replaced by the following weaker version: for
every CNPS (or CPS) $\mu _{i}$,%
\begin{equation*}
\bar{\rho}_{i}\left( \mu _{i}\right) :=\left\{ s_{i}\in S_{i}:\forall h\in
H_{i}\left( s_{i}\right) ,s_{i}\in \arg \max_{r_{i}\in S_{i}\left( h\right) }%
\mathbb{E}_{\mu _{i}}\left( U_{i}\left( r_{i},\cdot \right) |h\right)
\right\} \text{.}
\end{equation*}%
In words, $\bar{\rho}_{i}\left( \mu _{i}\right) $\ is the set of\textit{\ }%
\textbf{weak sequential best replies} to $\mu _{i}$: a strategy $s_{i}\in 
\bar{\rho}_{i}\left( \mu _{i}\right) $ must be optimal only at histories
that $s_{i}$ allows. Consequently, $\bar{\rho}_{i}\left( \cdot \right) $
does not distinguish between strategies in the same equivalence class; put
differently, weak sequential optimality determines reduced rather than full
strategies.

Redefine also iterated admissibility---i.e., the procedure $\left( \hat{S}%
^{n}\right) _{n\in \mathbb{N}}$---in terms of reduced strategies. Going
through the proofs in the Appendix, it is easy to see how the same proofs,
with minor changes, show that there are analogues of Theorems 1 and 2 in terms of
reduced\ strategies and weak sequential optimality.

\bigskip

\noindent \textbf{Acknowledgements}. I thank three anonymous referees,
Pierpaolo Battigalli and Emiliano Catonini for their comments and helpful
feedback. Needless to say, all mistakes are my own responsibility.

\section*{Appendix: Proofs of Theorem 1 and Theorem 2}

The proofs of Theorems 1 and 2 make use of the following properties of the standard
part---see Theorem 1.6.7 in Hurd and Loeb (1985). For any $x,y\in $ $^{\ast }%
\mathbb{R}$,

P1: $\mathrm{st}\left( x\pm y\right) =\mathrm{st}\left( x\right) \pm \mathrm{%
st}\left( y\right) $,

P2: $\mathrm{st}\left( xy\right) =\mathrm{st}\left( x\right) \mathrm{st}%
\left( y\right) $,

P3: $\mathrm{st}\left( y/x\right) =\frac{\mathrm{st}\left( y\right) }{%
\mathrm{st}\left( x\right) }$ if $\mathrm{st}\left( x\right) \neq 0$,

P4: $\mathrm{st}\left( x\right) \leq \mathrm{st}\left( y\right) $\ if $x\leq
y$.

\bigskip

In the remainder of this appendix, we fix a game $G:=\left\langle I,\bar{H}%
,\left( A_{i},u_{i}\right) _{i\in I}\right\rangle $. The following lemma
will be used in the proof of Theorem 1. As
usual, for all $s_{i}\in S_{i}$\ and $\nu _{i}\in $ $^{\ast }\Delta (S_{-i})$%
, let%
\begin{equation*}
U_{i}(s_{i},\nu _{i}):=\sum_{s_{-i}\in S_{-i}}U_{i}(s_{i},s_{-i})\nu
_{i}\left( s_{-i}\right) \text{.}
\end{equation*}

\begin{lemma}
\label{Lemma IA nonstandard}Fix $n\in \mathbb{N}$. Then, for every $s_{i}\in 
\hat{S}_{i}^{n}$ there exists $\nu _{i}\in $ $^{\ast }\Delta (S_{-i})$ such
that

(i) $\nu _{i}\left( s_{-i}\right) >0$ for every $s_{-i}\in S_{-i}$,

(ii) $s_{i}\in \mathrm{argmax}_{s_{i}^{\prime }\in S_{i}}U_{i}(s_{i}^{\prime
},\nu _{i})$, and

(iii) for every $m\in \left\{ 0,...,n-1\right\} $\ and for every $s_{-i}\in 
\hat{S}_{-i}^{m}$,%
\begin{equation*}
\mathrm{st}\left( \frac{\nu _{i}\left( S_{-i}\backslash \hat{S}%
_{-i}^{m}\right) }{\nu _{i}\left( s_{-i}\right) }\right) =0\text{.}
\end{equation*}
\end{lemma}

\noindent \textbf{Proof}. By Lemma 2 in Veronesi (1997) (see also Lemma E.1
in Brandenburger et al. 2008), for each $\ell =0,...,n-1$, there exists $\nu
_{\ell }\in \Delta (S_{-i})$ such that \textrm{supp}$\mathrm{\,}\nu _{\ell }=%
\hat{S}_{-i}^{n-1-\ell }$\ and $s_{i}\in $\textrm{argmax}$_{s_{i}^{\prime
}\in S_{i}}U_{i}(s_{i}^{\prime },\nu _{\ell })$. Pick any $\epsilon \in 
\mathrm{Hal}^{+}\left( 0\right) $\ such that $\epsilon >0$. A (non-standard)
measure $\nu _{i}\in $ $^{\ast }\Delta (S_{-i})$ is constructed as follows.
Define, for all $s_{-i}\in S_{-i}$,%
\begin{equation*}
\eta \left( s_{-i}\right) :=\nu _{0}\left( s_{-i}\right) +\sum_{\ell
=1}^{n-1}\epsilon ^{\ell }\nu _{\ell }\left( s_{-i}\right) \text{,}
\end{equation*}%
and%
\begin{equation*}
\nu _{i}\left( s_{-i}\right) :=\frac{\eta \left( s_{-i}\right) }{%
\sum_{s_{-i}^{\prime }\in S_{-i}}\eta \left( s_{-i}^{\prime }\right) }\text{%
.}
\end{equation*}%
Thus, $\nu _{i}(s_{-i})>0$ for each $s_{-i}\in S_{-i}$, and it can be easily
checked that $s_{i}\in $\textrm{argmax}$_{s_{i}^{\prime }\in
S_{i}}U_{i}(s_{i}^{\prime },\nu _{i})$. It remains to be shown that
Condition (iii) holds. By definition of $\nu _{i}$, it is enough to show
that, for all $m\leq n-1$\ and $s_{-i}\in \hat{S}_{-i}^{m}$,%
\begin{equation*}
\mathrm{st}\left( \frac{\sum_{s_{-i}^{\prime }\in S_{-i}\backslash \hat{S}%
_{-i}^{m}}\eta \left( s_{-i}^{\prime }\right) }{\eta \left( s_{-i}\right) }%
\right) =0\text{.}
\end{equation*}%
To this end, fix any $m\leq n-1$. Let $s_{-i}\in \hat{S}_{-i}^{m}$. Notice
that, if $s_{-i}\in \hat{S}_{-i}^{n-1}$, then $\mathrm{st}\left( \eta \left(
s_{-i}\right) \right) =\nu _{0}\left( s_{-i}\right) >0$. Thus, using P1-P3,%
\begin{equation*}
\mathrm{st}\left( \frac{\sum_{s_{-i}^{\prime }\in S_{-i}\backslash \hat{S}%
_{-i}^{m}}\eta \left( s_{-i}^{\prime }\right) }{\eta \left( s_{-i}\right) }%
\right) =\frac{\mathrm{st}\left( \sum_{\ell =1}^{n-1}\epsilon ^{\ell }\nu
_{\ell }\left( S_{-i}\backslash \hat{S}_{-i}^{m}\right) \right) }{\mathrm{st}%
\left( \eta \left( s_{-i}\right) \right) }=0\text{.}
\end{equation*}%
With this, consider the case when $s_{-i}\in \hat{S}_{-i}^{m}\backslash \hat{%
S}_{-i}^{n-1}$. If $s_{-i}\in \hat{S}_{-i}^{n-2}$ (which implies $m<n-1$),
then $\nu _{0}\left( s_{-i}\right) =0$, so that%
\begin{eqnarray*}
\eta \left( s_{-i}\right) &=&\epsilon \left[ \nu _{1}\left( s_{-i}\right) +%
\frac{\epsilon ^{2}}{\epsilon }\nu _{2}\left( s_{-i}\right) +...+\frac{%
\epsilon ^{n-1}}{\epsilon }\nu _{n-1}\left( s_{-i}\right) \right] \\
&=&\epsilon \left( \sum_{\ell =1}^{n-1}\frac{\epsilon ^{\ell }}{\epsilon }%
\nu _{\ell }\left( s_{-i}\right) \right) \text{.}
\end{eqnarray*}%
Hence,%
\begin{equation*}
\frac{\sum_{s_{-i}^{\prime }\in S_{-i}\backslash \hat{S}_{-i}^{m}}\eta
\left( s_{-i}^{\prime }\right) }{\eta \left( s_{-i}\right) }=\frac{%
\sum_{\ell =1}^{n-1}\epsilon ^{\ell }\nu _{\ell }\left( S_{-i}\backslash 
\hat{S}_{-i}^{m}\right) }{\epsilon }\cdot \frac{1}{\sum_{\ell =1}^{n-1}%
\frac{\epsilon ^{\ell }}{\epsilon }\nu _{\ell }\left( s_{-i}\right) }\text{.}
\end{equation*}%
It turns out that%
\begin{equation*}
\mathrm{st}\left( \frac{1}{\sum_{\ell =1}^{n-1}\frac{\epsilon ^{\ell }}{%
\epsilon }\nu _{\ell }\left( s_{-i}\right) }\right) =\frac{1}{\nu _{1}\left(
s_{-i}\right) }\text{,}
\end{equation*}%
and%
\begin{eqnarray*}
\mathrm{st}\left( \frac{\sum_{\ell =1}^{n-1}\epsilon ^{\ell }\nu _{\ell
}\left( S_{-i}\backslash \hat{S}_{-i}^{m}\right) }{\epsilon }\right) &=&%
\mathrm{st}\left( \frac{\sum_{\ell =2}^{n-1}\epsilon ^{\ell }\nu _{\ell
}\left( S_{-i}\backslash \hat{S}_{-i}^{m}\right) }{\epsilon }\right) \\
&=&\mathrm{st}\left( \sum_{\ell =2}^{n-1}\frac{\epsilon ^{\ell }}{\epsilon }%
\nu _{\ell }\left( S_{-i}\backslash \hat{S}_{-i}^{m}\right) \right) \\
&=&0\text{,}
\end{eqnarray*}%
where the second equality follows from the fact that $m<n-1$, which yields $%
\nu _{1}\left( S_{-i}\backslash \hat{S}_{-i}^{m}\right) =0$ (recall that 
\textrm{supp}$\mathrm{\,}\nu _{1}=\hat{S}_{-i}^{n-2}$). Hence,%
\begin{equation*}
\mathrm{st}\left( \frac{\sum_{s_{-i}^{\prime }\in S_{-i}\backslash \hat{S}%
_{-i}^{m}}\eta \left( s_{-i}^{\prime }\right) }{\eta \left( s_{-i}\right) }%
\right) =0\cdot \frac{1}{\nu _{1}\left( s_{-i}\right) }=0\text{.}
\end{equation*}%
The case when $s_{-i}\in \hat{S}_{-i}^{m}\backslash \hat{S}_{-i}^{n-2}$ and $%
s_{-i}\in \hat{S}_{-i}^{n-3}$ (which implies $m<n-2$) is similar: it turns
out that $\nu _{1}\left( s_{-i}\right) =0$ and%
\begin{eqnarray*}
\eta \left( s_{-i}\right) &=&\epsilon ^{2}\nu _{2}\left( s_{-i}\right)
+...+\epsilon ^{n-1}\nu _{n-1}\left( s_{-i}\right) \\
&=&\epsilon ^{2}\sum_{\ell =2}^{n-1}\frac{\epsilon ^{\ell }}{\epsilon ^{2}}%
\nu _{\ell }\left( s_{-i}\right) \text{.}
\end{eqnarray*}%
By noting that $m<n-2$, it is easily deduced that $\nu _{1}\left(
S_{-i}\backslash \hat{S}_{-i}^{m}\right) =\nu _{2}\left( S_{-i}\backslash 
\hat{S}_{-i}^{m}\right) =0$. Thus,%
\begin{equation*}
\mathrm{st}\left( \frac{\epsilon ^{3}\nu _{2}\left( S_{-i}\backslash \hat{S}%
_{-i}^{m}\right) +...+\epsilon ^{n}\nu _{n-1}\left( S_{-i}\backslash \hat{S}%
_{-i}^{m}\right) }{\epsilon ^{2}}\right) =0
\end{equation*}%
and%
\begin{equation*}
\mathrm{st}\left( \frac{1}{\sum_{\ell =2}^{n-1}\frac{\epsilon ^{\ell }}{%
\epsilon ^{2}}\nu _{\ell }\left( s_{-i}\right) }\right) =\frac{1}{\nu
_{2}\left( s_{-i}\right) }\text{,}
\end{equation*}%
which yields the result. Proceeding this way, the claim follows.\hfill $%
\blacksquare $

\bigskip

\noindent \textbf{Proof of Theorem 1}. By
induction on $n\in \mathbb{N}_{0}$.

(\textit{Basis step}) Immediate, since $S^{0}=\hat{S}^{0}=S$.

(\textit{Inductive step}) Suppose that the result is true for each $m\leq n$%
. We show that it is true for each $m\leq n+1$.

Fix a player $i$\ and $s_{i}\in S_{i}$. Arguing by contraposition, suppose
that $s_{i}\notin \hat{S}_{i}^{n+1}$. Since $s_{i}$\ is weakly dominated
with respect to $\hat{S}^{n-1}$, there exists $\sigma _{i}\in \Delta (\hat{S}%
_{i}^{n+1})$ such that $U_{i}(\sigma _{i},s_{-i})\geq U_{i}(s_{i},s_{-i})$\
for every $s_{-i}\in \hat{S}_{-i}^{n}$ and $U_{i}(\sigma _{i},s_{-i}^{\prime
})>U_{i}(s_{i},s_{-i}^{\prime })$\ for some $s_{-i}^{\prime }\in \hat{S}%
_{-i}^{n}$. By the inductive hypothesis, $\hat{S}_{-i}^{n}=S_{-i}^{n}$. Pick
any $\mu _{i}\in \cap _{l=0}^{n}\mathrm{cSB}_{i}(S_{-i}^{l})$. Since every $%
s_{-i}\in S_{-i}=S_{-i}\left( \varnothing \right) $\ is assigned strictly
positive probability by $\mu _{i}\left( \cdot \left\vert S_{-i}\right.
\right) $, it follows that, for all $s_{-i}\in S_{-i}^{n}$,%
\begin{equation*}
U_{i}(\sigma _{i},s_{-i})\mu _{i}\left( s_{-i}\left\vert S_{-i}\right.
\right) -U_{i}(s_{i},s_{-i})\mu _{i}\left( s_{-i}\left\vert S_{-i}\right.
\right) \geq 0
\end{equation*}%
and, for $s_{-i}^{\prime }\in S_{-i}^{n}$,%
\begin{equation*}
U_{i}(\sigma _{i},s_{-i})\mu _{i}\left( s_{-i}^{\prime }\left\vert
S_{-i}\right. \right) -U_{i}(s_{i},s_{-i}^{\prime })\mu _{i}\left(
s_{-i}^{\prime }\left\vert S_{-i}\right. \right) >0\text{.}
\end{equation*}%
Then%
\begin{equation*}
\sum_{s_{-i}\in S_{-i}^{n}}\left( U_{i}(\sigma _{i},s_{-i})\mu _{i}\left(
s_{-i}\left\vert S_{-i}\right. \right) -U_{i}(s_{i},s_{-i})\mu _{i}\left(
s_{-i}\left\vert S_{-i}\right. \right) \right) >0\text{,}
\end{equation*}%
which implies $U_{i}(s_{i},\mu _{i}\left( s_{-i}\left\vert S_{-i}\right.
\right) )<U_{i}(\sigma _{i},\mu _{i}\left( s_{-i}\left\vert S_{-i}\right.
\right) )$ for all $s_{-i}\in S_{-i}^{n}$. Hence, $s_{i}\notin \rho _{i}(\mu
_{i})$. (For, if $s_{i}\in \rho _{i}(\mu _{i})$, then $U_{i}(s_{i},\mu
_{i}\left( s_{-i}\left\vert S_{-i}\right. \right) )\geq U_{i}(\sigma
_{i},\mu _{i}\left( s_{-i}\left\vert S_{-i}\right. \right) )$ for all $%
\sigma _{i}\in \Delta (S_{i})$\ and $s_{-i}\in S_{-i}$.) Thus, $s_{i}\notin
S_{i}^{n+1}$. This argument shows that $S^{n+1}\subseteq \hat{S}^{n+1}$.

Conversely, let $s_{i}\in \hat{S}_{i}^{n+1}$. By Lemma 1, there exists $\nu _{i}\in $ $^{\ast }\Delta (S_{-i})$ such that

(i) $\nu _{i}\left( s_{-i}\right) >0$ for every $s_{-i}\in S_{-i}$,

(ii) $s_{i}\in \mathrm{argmax}_{s_{i}^{\prime }\in S_{i}}U_{i}(s_{i}^{\prime
},\nu _{i})$, and

(iii) for every $m\in \left\{ 0,...,n\right\} $\ and for every $s_{-i}\in 
\hat{S}_{-i}^{m}$,%
\begin{equation}
\mathrm{st}\left( \frac{\nu _{i}\left( S_{-i}\backslash \hat{S}%
_{-i}^{m}\right) }{\nu _{i}\left( s_{-i}\right) }\right) =0\text{.}
\label{standard part for PB}
\end{equation}%
With this, define a CNPS $\mu _{i}:=\left( \mu _{i}\left( \cdot \left\vert
S_{-i}\left( h\right) \right. \right) \right) _{h\in H}\in $ $^{\ast }\Delta
^{\mathcal{S}_{-i}}\left( S_{-i}\right) $\ by letting $\mu _{i}\left( \cdot
\left\vert S_{-i}\right. \right) :=\nu _{i}$, and, for each $h\neq
\varnothing $, the measure $\mu _{i}\left( \cdot \left\vert S_{-i}\left(
h\right) \right. \right) $ is defined by conditioning. Since $s_{i}\in 
\mathrm{argmax}_{s_{i}^{\prime }\in S_{i}}U_{i}(s_{i}^{\prime },\mu
_{i}\left( \cdot \left\vert S_{-i}\right. \right) )$\ and $\mu _{i}\left(
S_{-i}\left( h\right) \left\vert S_{-i}\right. \right) >0$\ for all $h\neq
\varnothing $, it follows from standard results (see, e.g., Battigalli et
al. 2024, Theorem 34) that $s_{i}\in \rho _{i}(\mu _{i})$. It remains to be
shown that, for each $m=1,...,n$, the event $S_{-i}^{m}$ is c-strongly
believed under $\mu _{i}$. By the inductive hypothesis, $S_{-i}^{m}=\hat{S}%
_{-i}^{m}$ for all $m=1,...,n$. So, consider any $S_{-i}^{m}$\ ($m\leq n$)
and $h\in H$\ such that $S_{-i}^{m}\cap S_{-i}\left( h\right) \neq \emptyset 
$. Then%
\begin{equation*}
\mu _{i}\left( \hat{S}_{-i}^{m}\left\vert S_{-i}\left( h\right) \right.
\right) =\frac{\nu _{i}\left( \hat{S}_{-i}^{m}\cap S_{-i}\left( h\right)
\right) }{\nu _{i}\left( S_{-i}\left( h\right) \right) }\text{.}
\end{equation*}%
Hence, for all $s_{-i}\in S_{-i}^{m}\cap S_{-i}\left( h\right) $,%
\begin{equation*}
\mathrm{st}\left( \frac{\mu \left( S_{-i}\backslash \hat{S}%
_{-i}^{m}\left\vert S_{-i}\left( h\right) \right. \right) }{\mu \left(
s_{-i}\left\vert S_{-i}\left( h\right) \right. \right) }\right) =\mathrm{st}%
\left( \frac{\nu _{i}\left( \left( S_{-i}\backslash \hat{S}_{-i}^{m}\right)
\cap S_{-i}\left( h\right) \right) }{\nu _{i}\left( s_{-i}\right) }\right) =0%
\text{,}
\end{equation*}%
where the last equality follows from (6.1) and P4.
Thus, any $S_{-i}^{m}$\ ($m\leq n$)\ is c-strongly believed under $\mu _{i}$%
, as required.\hfill $\blacksquare $

\bigskip

\noindent \textbf{Proof of Theorem 2}. We show,
by induction on $n\in \mathbb{N}_{0}$, that $\hat{S}^{n}=\bar{S}^{n}$. With
this, Theorem 1 yields the result.

(\textit{Basis step}) Immediate, since $\hat{S}^{0}=\bar{S}^{0}=S$.

(\textit{Inductive step}) Suppose that the result is true for each $m\leq n$%
. We show that it is true for each $m\leq n+1$.

Let $s_{i}\in \hat{S}_{i}^{n+1}$. Then $s_{i}\in \hat{S}_{i}^{n}$, and so,
by the inductive hypothesis, $s_{i}\in \bar{S}_{i}^{n}$. So, $s_{i}\in \rho
_{i}(\mu _{i})$\ for some CPS $\mu _{i}\in \Delta ^{\mathcal{S}_{-i}}\left(
S_{-i}\right) $\ such that, for every $h\in H$,%
\begin{equation*}
\bar{S}_{-i}^{n-1}\cap S_{-i}\left( h\right) \neq \emptyset \Rightarrow 
\mathrm{supp}\mu _{i}\left( \cdot |S_{-i}\left( h\right) \right) =\bar{S}%
_{-i}^{n-1}\cap S_{-i}\left( h\right) \text{.}
\end{equation*}%
By Lemma E.1 in Brandenburger et al. (2008) (or Lemma 2 in Veronesi 1997),
there exists $\nu \in \Delta (S_{-i})$ such that \textrm{supp}$\mathrm{\,}%
\nu =\hat{S}_{-i}^{n}=\bar{S}_{-i}^{n}$\ and $s_{i}\in $\textrm{argmax}$%
_{s_{i}^{\prime }\in S_{i}}U_{i}(s_{i}^{\prime },\nu )$. Define a CPS $\hat{%
\mu}_{i}\in \Delta ^{\mathcal{S}_{-i}}\left( S_{-i}\right) $\ as follows:
for each $h\in H$,%
\begin{equation*}
\hat{\mu}_{i}\left( \cdot \left\vert S_{-i}\left( h\right) \right. \right)
:=\left\{ 
\begin{tabular}{ll}
$\frac{\nu \left( \cdot \cap S_{-i}\left( h\right) \right) }{\nu \left(
S_{-i}\left( h\right) \right) }$, & if $\bar{S}_{-i}^{n}\cap S_{-i}\left(
h\right) \neq \emptyset $, \\ 
$\mu _{i}\left( \cdot \left\vert S_{-i}\left( h\right) \right. \right) $, & 
otherwise.%
\end{tabular}%
\right. 
\end{equation*}%
It is immediate to check that $\hat{\mu}_{i}$\ satisfies the required
properties and $s_{i}\in \rho _{i}(\hat{\mu}_{i})$.

For the converse, pick any $s_{i}\in \bar{S}_{i}^{n+1}$. Then, by the
inductive hypothesis, $s_{i}\in \hat{S}_{i}^{m}$ for all $m\leq n$. It must
be shown that $s_{i}\in \hat{S}_{i}^{n+1}$. Since $s_{i}\in \bar{S}_{i}^{n+1}
$, there exists $\mu _{i}\in \Delta ^{\mathcal{S}_{-i}}\left( S_{-i}\right) $%
\ such that $\mathrm{supp}\mu _{i}\left( \cdot |S_{-i}\right) =\bar{S}%
_{-i}^{n}=\hat{S}_{i}^{n}$ and $s_{i}\in \rho _{i}\left( \mu _{i}\right) $.
In particular, $s_{i}\in $\textrm{argmax}$_{s_{i}^{\prime }\in
S_{i}}U_{i}(s_{i}^{\prime },\mu _{i}\left( \cdot |S_{-i}\right) )$. By Lemma
4 in Pearce (1984), $s_{i}$\ is admissible with respect to $\hat{S}^{n}$.
Hence, $s_{i}\in \hat{S}_{i}^{n+1}$.\hfill $\blacksquare $

\nocite{*}
\bibliographystyle{eptcs}
\bibliography{mainpaper}
\end{document}